\documentclass{aa}
\usepackage{natbib}
\usepackage{siunitx}
 \DeclareSIUnit\gauss{gauss}
\usepackage{amsmath}
\usepackage{nccmath}
\usepackage{graphicx }

 \usepackage{natbib,twoopt}
\usepackage[breaklinks=true]{hyperref} 
\bibpunct{(}{)}{;}{a}{}{,} 
\makeatletter
\newcommandtwoopt{\citeads}[3][][]{\href{http://adsabs.harvard.edu/abs/#3}%
{\def\hyper@linkstart##1##2{}%
\let\hyper@linkend\@empty\citealp[#1][#2]{#3}}}
\newcommandtwoopt{\citepads}[3][][]{\href{http://adsabs.harvard.edu/abs/#3}%
{\def\hyper@linkstart##1##2{}%
\let\hyper@linkend\@empty\citep[#1][#2]{#3}}}
\newcommandtwoopt{\citetads}[3][][]{\href{http://adsabs.harvard.edu/abs/#3}%
{\def\hyper@linkstart##1##2{}%
\let\hyper@linkend\@empty\citet[#1][#2]{#3}}}
\newcommandtwoopt{\citeyearads}[3][][]%
{\href{http://adsabs.harvard.edu/abs/#3}
{\def\hyper@linkstart##1##2{}%
\let\hyper@linkend\@empty\citeyear[#1][#2]{#3}}}
\makeatother

\begin{document}

\title{Characterization of the umbra--penumbra boundary by the vertical component of the magnetic field}
\subtitle{Analysis of ground-based data from the GREGOR Infrared Spectrograph}

\author{P. Lindner\and R. Schlichenmaier \and N. Bello Gonz\'alez}

\titlerunning{Characterization of the umbra--penumbra boundary by $B_\perp$}
\authorrunning{Lindner et al.} 

\institute{Leibniz-Institut für Sonnenphysik (KIS)}

\abstract{The vertical component of the magnetic field was found to reach a constant value at the boundary between penumbra and umbra of stable sunspots in a recent statistical study of Hinode/SP data. This finding has profound implications as it can serve as a criterion to distinguish between fundamentally different magneto-convective modes operating in the sun.}
{The objective of this work is to verify the existence of a constant value for the vertical component of the magnetic field ($B_\perp$) at the boundary between umbra and penumbra from ground-based data in the near-infrared wavelengths and to determine its value for the GREGOR Infrared Spectrograph (GRIS@GREGOR) data. This is the first statistical study on the  {Jur{\v{c}}{\'a}k criterion} with ground-based data, and we compare it with the results from space-based data (Hinode/SP and SDO/HMI).}
{Eleven spectropolarimetric data sets from the GRIS@GREGOR slit-spectograph containing fully-fledged stable sunspots were selected from the GRIS archive (\href{sdc.leibniz-kis.de}{sdc.leibniz-kis.de}). SIR inversions including a polarimetric straylight correction are used to produce maps of the magnetic field vector using the Fe I \SI{15648}{\angstrom} and \SI{15662}{\angstrom} lines. Averages of $B_\perp$ along the contours between penumbra and umbra are analyzed for the 11 data sets. In addition, contours at the resulting $B_\perp^{\rm const}$ are drawn onto maps and compared to intensity contours. The geometric difference between these contours, $\Delta P$, is calculated for each data set.}
{Averaged over the 11 sunspots, we find a value of $B_\perp^{\rm const} =\SI[separate-uncertainty=true]{1787(100)}{\gauss}$. The difference from the values previously derived from Hinode/SP and SDO/HMI data is explained by instrumental differences and by the formation characteristics of the respective lines that were used. Contours at $B_\perp = B_\perp^{\rm const}$ and contours calculated in intensity maps match from a visual inspection and the geometric distance  $\Delta P$ was found to be on the order of 2 pixels. Furthermore, the standard deviation between different data sets of averages along umbra--penumbra contours is smaller for $B_\perp$ than for $B_\parallel$ by a factor of 2.4.}{Our results provide further support to the {Jur{\v{c}}{\'a}k criterion} with the existence of an invariable value $B_\perp^{\rm const}$ at the umbra--penumbra boundary. This fundamental property of sunspots can act as a constraining parameter in the calibration of analysis techniques that calculate magnetic fields. It also serves as a requirement for numerical simulations to be realistic. Furthermore, it is found that the geometric difference, $\Delta P$, between intensity contours and contours at $B_\perp=B_\perp^{\rm const}$ acts as an index of stability for sunspots.} 

\maketitle


\section{Introduction}
Sunspots are commonly divided into a central umbra and the surrounding penumbra. In the umbra, magnetic field strength values of up to \SI{3000}{\gauss} and more are observed and the field is mostly vertical. This leads to an effective hindering of convection. Although the observation of umbral substructures (e.g., umbral dots \citepads{2010ApJ...713.1282O}) {
 show that convection is not totally suppressed, heat flux values in the umbra are reduced to \SIrange[range-units=single,range-phrase = -]{5}{25}{\percent} of the quiet-sun value \citepads{2011LRSP....8....4B}. In the penumbra, magnetic fields are observed to be weaker than in the umbra and more inclined  \citepads{2011LRSP....8....3R}, forming a funnel-like topology of the magnetic field. Penumbral heat flux values between those of umbra and quiet-sun are found, which indicates that convection is hindered less effectively. Radially aligned filaments are observed in the intensity parameter and in the magnetic field vector parameters (see \citetads{2013A&A...557A..25T} for a detailed description of their magnetic topology). This study also describes two basic features of the penumbra: its uncombed structure, i.e., the magnetic field consists of two interlaced components \citepads{1993A&A...275..283S}, and the  outward-directed horizontal Evershed flow \citepads{1909Obs....32..291E}. \\
Despite  the structural differences listed above, the boundary between the penumbra and umbra of sunspots was, until recently, only defined by a threshold in intensity in solar images. Contours outlining the umbra were drawn on specific intensity level values, which depend  on several factors including the wavelength, spatial resolution, optical properties of the instrument, and seeing conditions. \citetads{2018A&A...611L...4J}, however, found that this boundary can also be defined by a constant value of the vertical component of the magnetic field (hereafter $B_\perp$) from a statistical study. Before this statistical evidence, the existence of this constant was suggested for Hinode/SP \citepads{2011A&A...531A.118J} and VTT/GFPI data \citepads{2015A&A...580L...1J}. The authors propose this constant value (estimated as \SI{1867}{\gauss} in the study of Hinode/SP data) to be the magnetic threshold limiting the operation of the magneto-convection type of the penumbra. This criterion is called the {Jur{\v{c}}{\'a}k criterion}. \citetads{2018A&A...620A.104S} then demonstrated, using SDO/HMI data, that a constant value of $B_\perp$ also outlines the umbra of a stable sunspot over a time span of several days. However, a different value for this constant of  \SI{1693}{\gauss} was found. In this study we investigate the existence of a constant value of $B_\perp$ for stable sunspots with ground-based telescope data for the first time and calculate its value, aiming to understand the different values from different data sources.

\section{Methods}
We produced maps of the magnetic field vector with the SIR inversion code \citepads{1992ApJ...398..375R} using 11 spectropolarimetric data sets from the GREGOR Infrared
Spectrograph (GRIS@GREGOR)  \citepads{2012AN....333..872C}.

\subsection{Data selection}
For this study, 11 data sets of fully-fledged sunspots were selected from the GRIS archive \footnote{\href{sdc.leibniz-kis.de}{sdc.leibniz-kis.de}},
 where observations since 2014 are publicly available. The GRIS instrument \citepads{2012AN....333..872C} is a slit-spectograph in the infrared regime providing high resolution spectropolarimetric data. It is attached to the 1.5 m GREGOR telescope situated at the Observatorio del Teide, Tenerife \citepads{2012AN....333..796S}; the telescope is equipped with an adaptive optics system \citepads{2012AN....333..863B}. \\
The selection criteria used to obtain our samples of sunspots were that a) the Fe I \SI{15648}{\angstrom} and the \SI{15662}{\angstrom} line were recorded simultaneously, b) the sunspot contained a fully-fledged penumbra, c)  the heliocentric angle was not larger than \SI{35}{\degree}, and d)   the data quality (especially seeing) was sufficient. In Table \ref{dataset_list}, a list of these data sets is shown. The recorded spectral windows were approximately \SI{40}{\angstrom} wide, with central positions that differed only slightly. The wavelength pixel size was $\delta_{\lambda} \approx \SI{0.04}{\angstrom}$  and the spatial pixel size, as obtained from a correlation with HMI data \citepads{2016A&A...596A...2B}, was $\delta_{xy} \approx \SI{0.135}{\arcsec}$. The data was demodulated and corrected with the standard GRIS pipelines \citepads[see][for a description of some of the included methods]{2016A&A...596A...4F} and we worked with the data cubes from the level 2 data, which includes a wavelength calibration and the derivation of the spot coordinates based on HMI images. In addition, we corrected for the daily variations of solar flux (creating an intensity gradient for subsequent slit positions) with a linear fit over the scanning direction.

\subsection{Straylight correction}
\label{straysec}
The SIR inversion code \citepads{1992ApJ...398..375R} offers the possibility of providing a spatial straylight profile $\vec{I_{\mathrm{stray}}}$ for each pixel:
\begin{ceqn}
\begin{align}
        \vec{I_{\mathrm{obs}}}=(1-\alpha)\vec{I^*}+\alpha \vec{I_{\mathrm{stray}}}
\label{sir_stray}
\end{align}
\end{ceqn}
Using Eq. \ref{sir_stray}, this straylight profile is added to the full-Stokes synthetic profile $\vec{I^*}$ of the respective atmosphere before the fit is compared to the actual observed full-Stokes profile $\vec{I_{\mathrm{obs}}}$. We calculated a specific full-Stokes spatial straylight profile for each pixel separately. Following the idea by \citetads{2016A&A...596A...2B}, we calculated the straylight profile as the sum over all pixels of the map. However, each pixel was weighted with a two-dimensional Gaussian centered at the respective pixel, so that neighboring pixels contribute more. The full width at half maximum (FWHM) of this Gaussian was \SI{47.1}{\arcsec} (equivalent to a standard deviation of $\sigma= \SI{20}{\arcsec}$), as estimated by \citetads{2016A&A...596A...2B}. The advantage of  a local straylight profile is that, in addition to intensity straylight profiles, meaningful Stokes {\em Q, U}, and  {\em V} straylight profiles can be obtained. Neighboring pixels are assumed to have a more similar polarization signal than far-away pixels, and therefore cancellation effects in  {\em Q, U}, and  {\em V} are less probable. \\
The factor $\alpha$ in Eq. \ref{sir_stray} controls how much of the actually observed profile is attributed to straylight. We calculated this straylight factor separately for each data set (sunspot map) by calculating the Stokes-I spectrum of the resulting effective intensity profile $\vec{I^*}$. Since our data sets are all at small heliocentric angles, we assumed that there are pixels in the umbra, where the magnetic field is parallel to the line of sight (LOS). For these pixels the magnetic fields are strong enough so that the Fe I line at \SI{15648}{\angstrom} with an effective Land\'e factor of 3 is fully split. Therefore, only the $\sigma$-components are expected to be seen and no $\pi$-component should be visible in the Stokes-I spectrum. Separately for each data set, the $\alpha$ value was increased in steps of 0.01 until the maximum Stokes-{\em I} value of $\vec{I^*}$    between the $\sigma$-components met the continuum intensity level. The condition we used was that less than 0.1 \% of the umbral pixels (depending on the data set, there are between 5000 and 45000 umbral pixels) were allowed to have a higher value than the continuum. This was done to ensure the stability of our computations in the cases where single pixels show irregular behavior. Actual straylight contamination might still be higher, but using higher values for $\alpha$ would lead to an undesired overcorrection (line center going into emission; see Fig. \ref{fig_alpha}) for numerous pixels.  This method of correcting for straylight can therefore be regarded as conservative. It is also simple enough that it can be applied automatically to many data sets, and no manual interaction is needed. The   average profile shown in Fig. \ref{fig_alpha} illustrates the effect of changing $\alpha$. The $\alpha$ value we used for data set 26apr14.007 is $0.10$. At this value, single pixels (i.e., 0.1 \% of the umbral pixels) are already in emission, while the average corrected profile is not yet in emission. \\ 
\begin{figure}[h]
\centering
   \resizebox{\hsize}{!}{\includegraphics{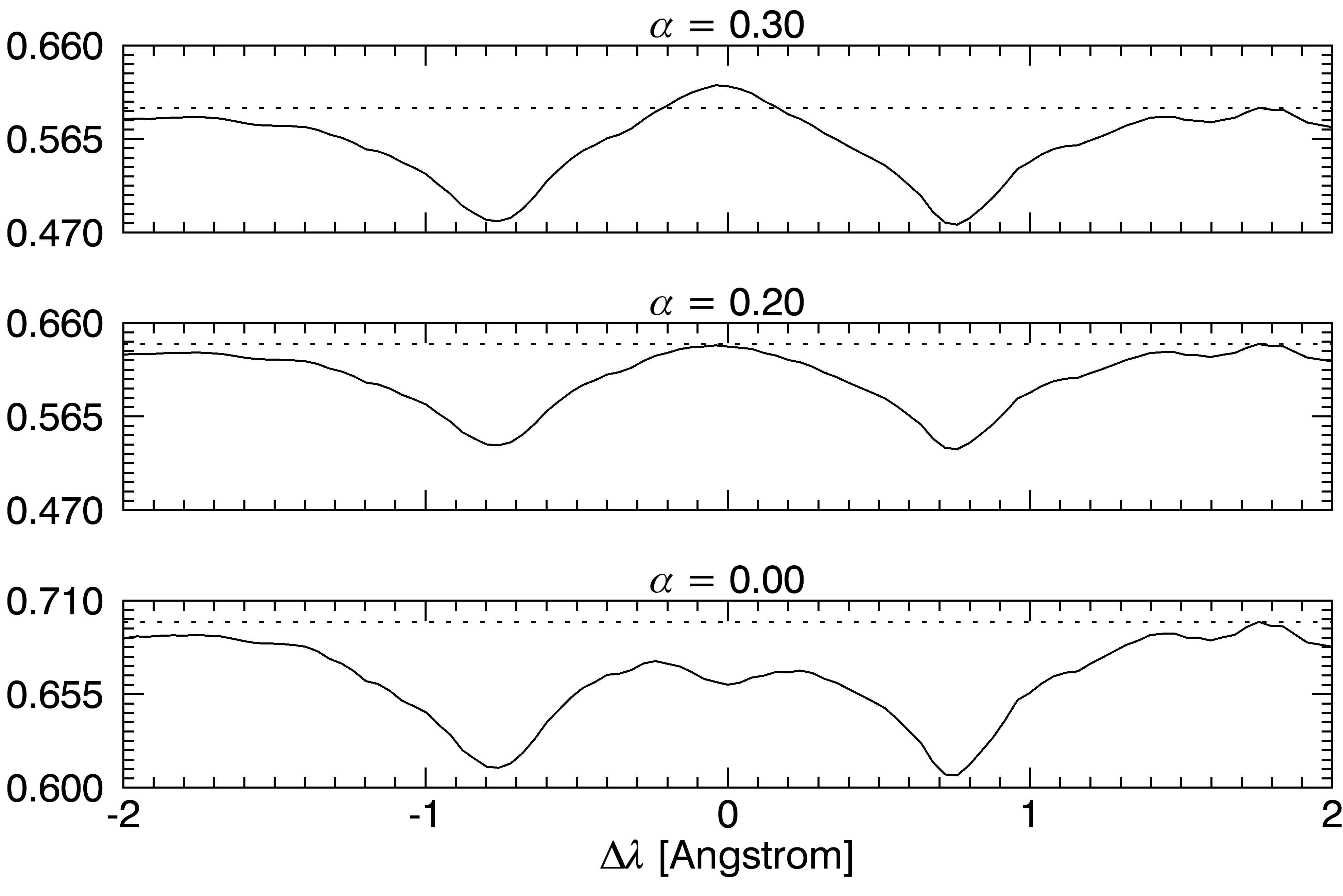}}
     \caption{Stokes-I spectra of straylight corrected profiles $\vec{I^*}$ for different straylight scaling factors $\alpha$ of the fully split \SI{15648}{\angstrom} line. Shown here is an average over selected umbral pixels where the magnetic field is assumed to be parallel to the LOS, to demonstrate the principle of the correction method. The lower graph shows the uncorrected profile where a central lobe is still present. The graph in the middle shows the corrected case where the intensity level at the line center meets the continuum level. The uppermost graph shows an example of overcorrection.} 
     \label{fig_alpha}
\end{figure}
The resulting values for $\alpha$ were mostly in the range between 0.10 and 0.15 (see Fig. \ref{contour_levels}) with one exception. The big sunspot of data set 21may16.003 had molecular blends in the dark parts of the umbra (present because of strong magnetic fields and reduced temperatures). Therefore, a calculation of $\alpha$ including a comparison between continuum intensity and intensity at the $\pi$-component (line center) was not possible in the umbra. For this special case, we used the average $\alpha$ value of the other ten data sets ($\overline{\alpha}=0.12$).

\subsection{Inversions}
The  11 sunspot spectropolarimetric data sets were inverted with the SIR (Stokes Inversion based on Response functions) code \citepads{1992ApJ...398..375R}. The Fe I \SI{15648}{\angstrom} and \SI{15662}{\angstrom} lines were inverted simultaneously. Each inversion run consisted of three cycles with (1,2,4) nodes for the temperature and (1,1,1) nodes for the magnetic field components and the LOS velocity. In the first cycle, equal weights for the Stokes parameters were used and in the last two, more weight was put on {\em Q, U}, and {\em V}. This inversion run was repeated three times,  each time randomizing the initial values for temperature, magnetic field strength, inclination, azimuth, and LOS velocity. The run with the best $\chi^2$-value (sum of the squared differences between observed and fitted profile, scaled with the signal-to-noise ratio) was chosen. In this  way a good compromise between computation time and performance in terms of stability, agreement of fit with input profiles, and smoothness of resulting maps was achieved. The initial model was based on the {hot umbra model} introduced by \citetads{1994A&A...291..622C}. However, the input model was randomized for each run to prevent the algorithm from finding a local minimum close to the initial model instead of finding the global minimum. \\
The magnetic field vector produced by the inversion was transformed from the LOS reference frame to the local reference frame (LRF) and the azimuthal ambiguity was resolved with the interactive AZAM code \citepads{1995ApJ...446..877L}. After that, maps of $B_\perp$ and $B_\parallel$ (the component of the magnetic field that is vertical and parallel to the local surface, respectively) were calculated as $B_\perp=B \cdot \sin(\theta)$ and $B_\parallel=B \cdot \cos(\theta)$ . An example of such a map is shown in Fig. \ref{examplemaps}.

\begin{figure}[h]
\centering
        \resizebox{\hsize}{!}{\includegraphics{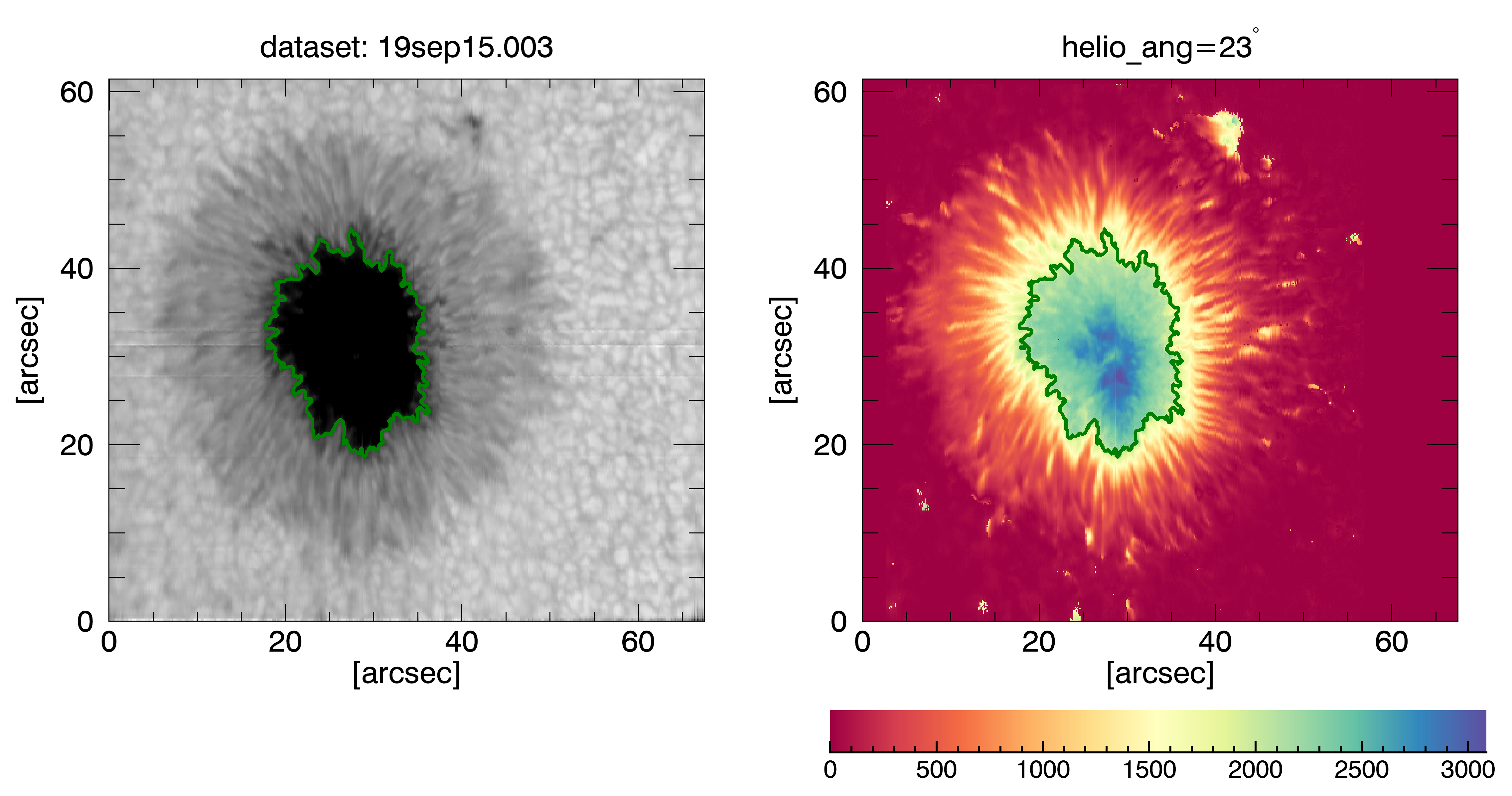}}
     \caption{Left: Intensity map with contour (in green) at the manually found level. Right: Map of the vertical component of the magnetic field, $B_\perp$, calculated from inversion results. The contour calculated from the intensity map is also drawn on the $B_\perp$  map in green.}
     \label{examplemaps}
\end{figure}

\subsection{Finding intensity contours}
The boundary between the penumbra and umbra was first defined as a contour at a specific value in the straylight-corrected continuum intensity maps. For space-based data like Hinode, a fixed value can be used to find this contour, like done e.g. by \citetads{2018A&A...611L...4J}. In order to investigate whether this is possible for GRIS data, the contour level value was at first chosen manually by increasing/decreasing the value until the resulting contour was considered well outlining the umbra from a visual inspection. An example of such a contour is shown in Fig. \ref{examplemaps}. 

\begin{table}
\caption{List of GRIS@GREGOR data sets that were used in this study, the time of the first slit position (UT), the heliocentric angle at the center of the field of view (FOV) ($\gamma$), the values of the straylight scaling factor ($\alpha$), manually found intensity value at which the contour between penumbra and umbra is calculated  ($I_\circ$, rounded numbers). This data can be accessed openly via \href{sdc.leibniz-kis.de}{sdc.leibniz-kis.de}}
\label{dataset_list}
\centering 
\begin{tabular}{l l c l l }
data set & UT time & $\gamma $ & $\alpha$ & $I_\circ$ \\
\hline \hline 
23apr15.003 & 09:46:36 & 5$^\circ$ & $0.03$       & $0.76$\\
12sep15.009 & 09:34:50 & 29$^\circ$   & $0.08$       & $0.76$\\
26apr14.007 & 10:43:44 & 20$^\circ$ & $0.10$       & $0.80$\\
03may14.012 & 14:05:02 & 6$^\circ$  & $0.11$      & $0.79$\\
21may16.003 & 14:44:58 & 17$^\circ$ & $0.12$ \tablefootmark{*}       & $0.72$\\
03sep17.008 & 10:29:12 & 17$^\circ$ & $0.13$       & $0.77$\\
29aug16.006 & 09:25:29 & 15$^\circ$ & $0.13$       & $0.78$\\
15sep15.001 & 08:10:06 & 34$^\circ$ & $0.14$       & $0.78$\\
19sep15.003 & 09:25:42 & 23$^\circ$ & $0.15$       & $0.71$\\
02sep17.004 & 09:07:23 & 25$^\circ$ & $0.16$       & $0.77$\\
16aug16.001 & 08:26:18 & 33$^\circ$ & $0.18$        & $0.78$\\
\hline
\end{tabular}
\tablefoot{
\tablefoottext{*}{$\alpha$ was inferred from other data sets (see Sect. \ref{straysec})}
}
 \end{table}

\begin{figure}[h]
\centering
        \resizebox{\hsize}{!}{\includegraphics{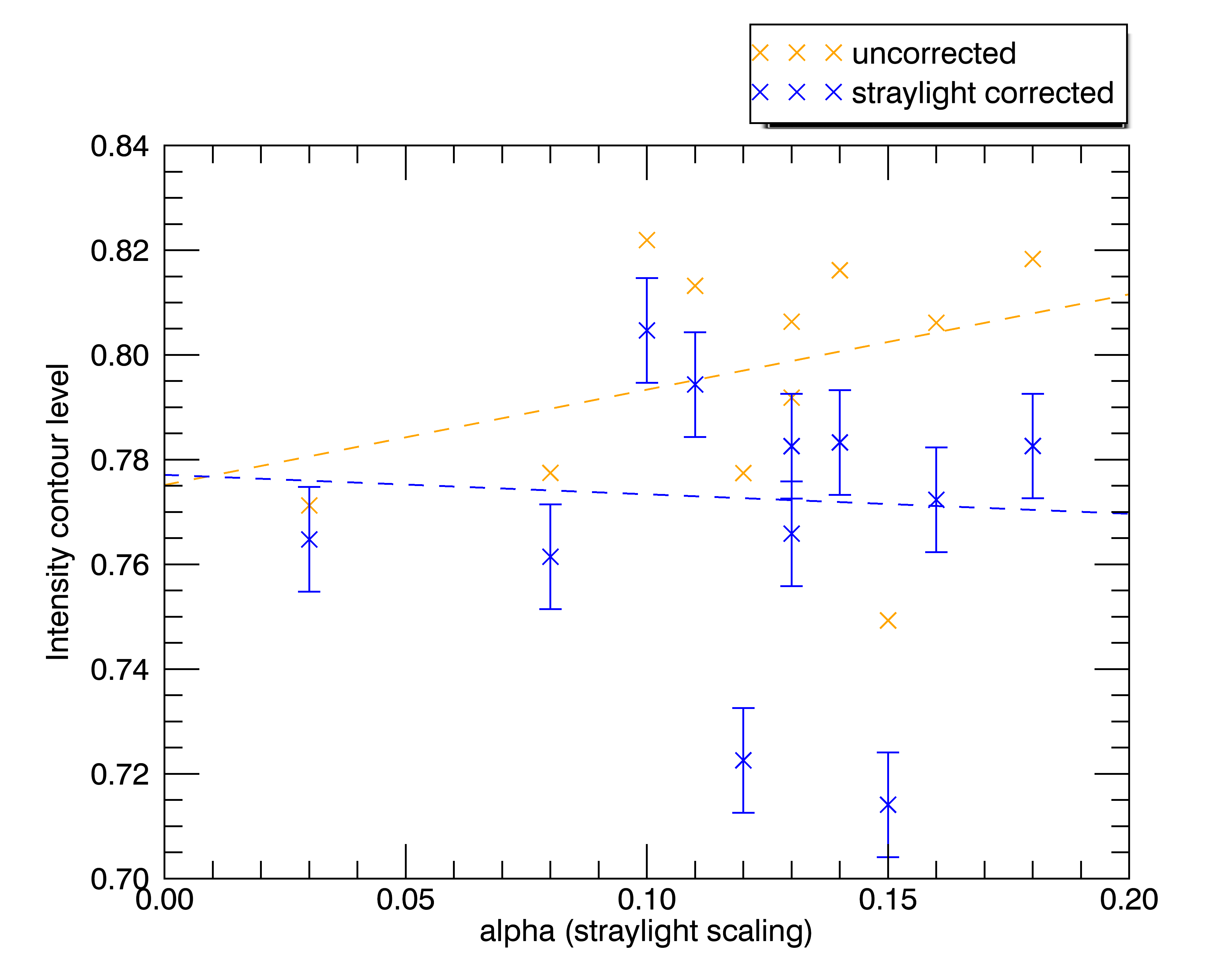}}
     \caption{Manually found intensity level values   for which satisfying contours were obtained. The blue data points represent the intensity contour values calculated in the straylight corrected intensity maps and the orange data points represent the values calculated in the uncorrected intensity maps. On the horizontal axis the straylight scaling factor $\alpha$ is depicted.}
     \label{contour_levels}
\end{figure}

The resulting values for each data set are shown in Table \ref{dataset_list} and are plotted in Fig. \ref{contour_levels}. There was a  difference of up to 0.09 between the values of the different data sets. However, by visual inspection we find that a change in the intensity level of more than 0.01  leads to contours that are clearly not at the umbra--penumbra boundary. This value of 0.01 also defines the error bars in Fig. \ref{contour_levels}. We therefore continued working with the manually found contours instead of using a fixed value. We ascribe the differences in contour level values mostly to different observing conditions, particularly  to {seeing}, that alter the contrast in the intensity maps. Our straylight correction is too simple and crude to remove all the effects of seeing. In order to verify the effect of our straylight correction, we also plotted intensity level values calculated in uncorrected maps (orange data points in Fig. \ref{contour_levels}). A linear regression (slope: 0.182) showed that in the uncorrected data there was a trend towards higher $\alpha$ values being related to higher values for the intensity contour. A linear regression with the straylight corrected data (slope: -0.037) showed that while the scattering of the data remained, this trend was removed by the straylight correction.

  
\section{Analysis}
The values of the vertical component of the magnetic field, $B_\perp$, were evaluated at the manually found intensity contours. Figure \ref{alongcontour} shows $B_\perp$ and $B_\parallel$ (vertical and horizontal component of the magnetic field, respectively) and $B_\mathrm{abs}$ (absolute value of the magnetic field) running along the intensity contour for one  data set.  Both $B_\perp$ and $B_\parallel$ show some outliers for most of the data sets, but $B_\parallel$ also shows slopes ranging over large parts of the contour.

\begin{figure}[h]
\centering
        \resizebox{\hsize}{!}{\includegraphics{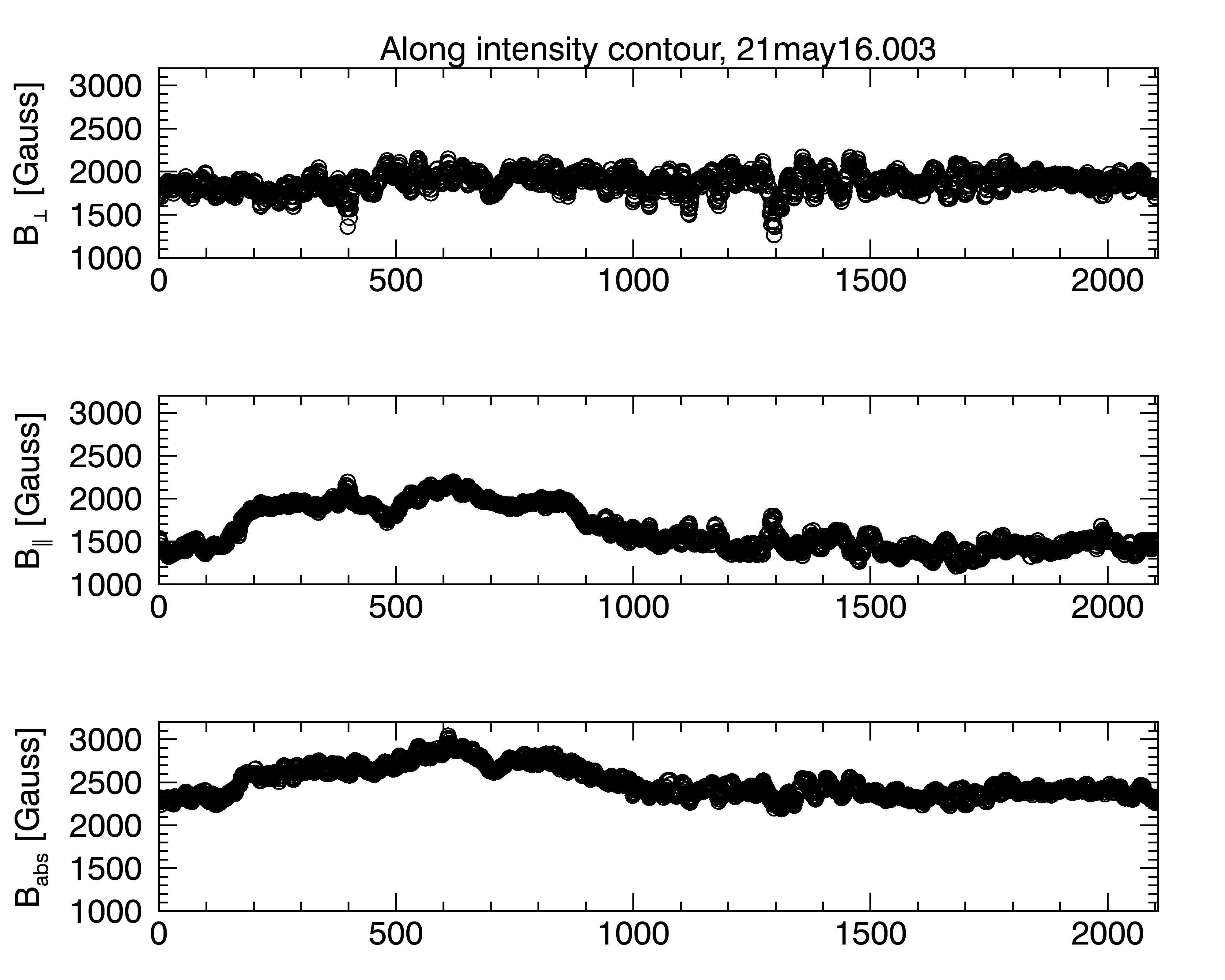}}
     \caption{$B_\perp$, $B_\parallel$, and $B_\mathrm{abs}$ running along the pixels of the intensity contour for one  data set. We note that $B_\perp$ and $B_\parallel$ are calculated from $B_\mathrm{abs}$ and the inclination $\theta$, so they are subject to errors from both $B_\mathrm{abs}$ and $\theta$. A direct comparison to $B_\mathrm{abs}$ is therefore difficult.}
     \label{alongcontour}
\end{figure}

\subsection{Averages over contours}
\label{bversec}
For each data set, the median over the values of $B_\perp$ and $B_\parallel$ along the intensity contour was calculated. We denote these average quantities by $\overline{B_\perp}$ and  $\overline{B_\parallel}$. Figure \ref{bversort} shows these values and how they are grouped around the average value between the data sets. We identify the average value of  $\overline{B_\perp}=\SI{1787}{\gauss}$ as the critical constant below which penumbral magneto-convection can operate. \\
 In order to compare the behavior of the averaged vertical and horizontal components of the magnetic field at the umbral boundary, i.e.,  $\overline{B_\perp}$ to $\overline{B_\parallel}$, Fig. \ref{bverrelsort} also shows these quantities normalized to their respective average values over the different data sets. From this visualization it becomes apparent that the  $\overline{B_\perp}$ values are less widely spread than the $\overline{B_\parallel}$ values.  The standard deviations of $\overline{B_\perp}$ and  $\overline{B_\parallel}$ with respect to the different data sets were $\sigma_\perp=0.052$ and $\sigma_\parallel=0.126$, normalized to the average value. \\

\begin{figure}[h]
\centering
        \resizebox{\hsize}{!}{\includegraphics{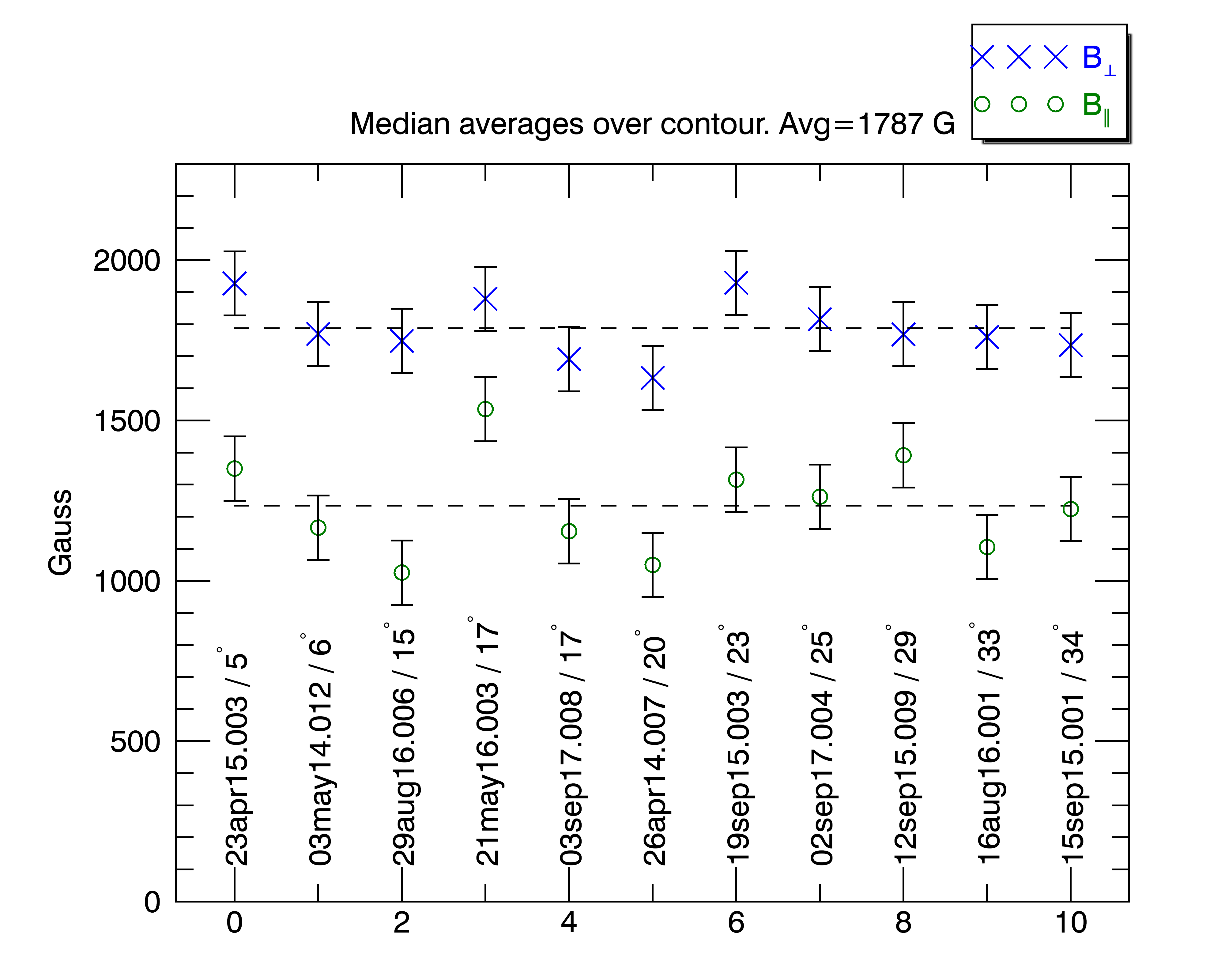}}
     \caption{$\overline{B_\perp}$ and  $\overline{B_\parallel}$ (median values over the intensity contour) for different data sets. The heliocentric angle is given with the data set titles. The dashed lines represent the value of $B_\perp$ and $B_\parallel$ averaged over all 11 maps.}
     \label{bversort}
\end{figure}

\begin{figure}[h]
\centering
        \resizebox{\hsize}{!}{\includegraphics{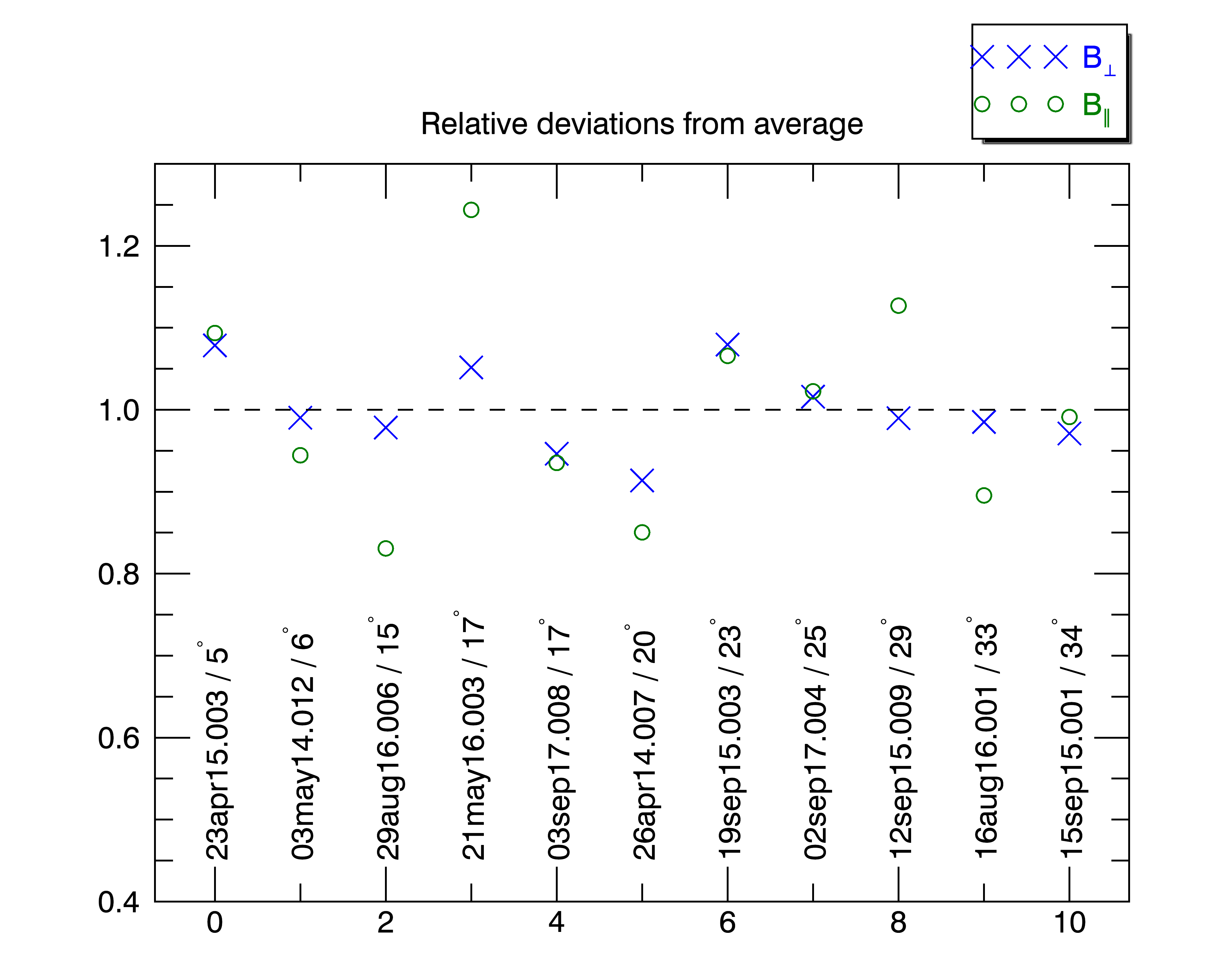}}
     \caption{$\overline{B_\perp}$ and  $\overline{B_\parallel}$ normalized to the averages over all data sets to allow for a direct comparison.}
     \label{bverrelsort}
\end{figure}

\subsection{Error estimation}
\label{intcont}
Errors in our calculations come from the errors in the $B_\perp$ and $B_\parallel$ maps and from errors in finding the intensity contours. As these contours were found by choosing the corresponding level values $I_c$ manually and from a visual inspection, they introduce a certain subjectivity. The experience from plotting contours for many (50+) GRIS data sets showed that contours at $I_c \pm 0.01$ could be chosen by a different person or in a second run. We denote the corresponding contours with $C_+$ and $C_-$. In Fig. \ref{varycontour}, these contours are plotted for two example data sets.
\begin{figure}[h]
\centering
        \resizebox{\hsize}{!}{\includegraphics{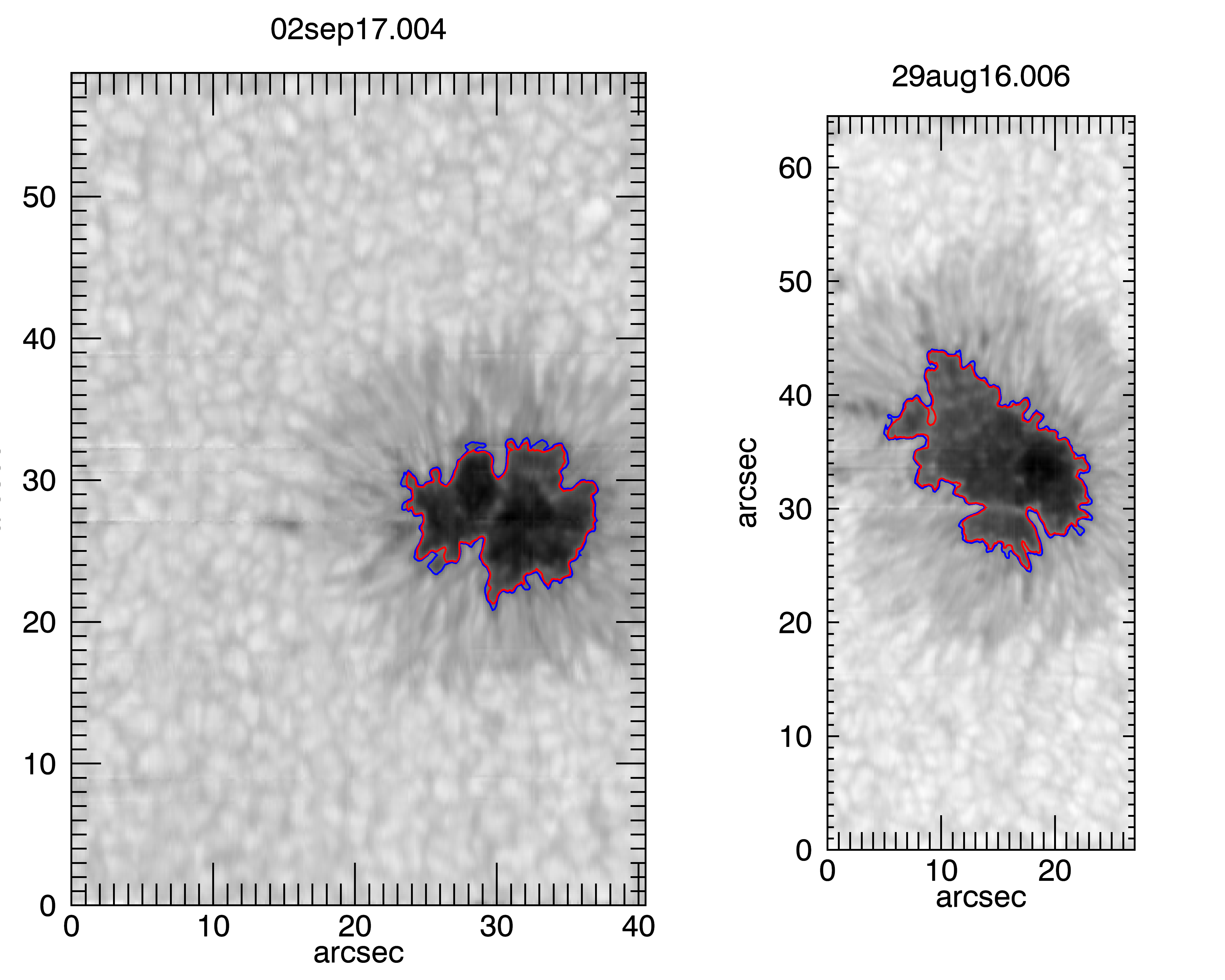}}
     \caption{Intensity contours with level values of $I_c + 0.01$ in blue and $I_c - 0.01$ in red. $I_c$ is the value that was used for the analysis.}
     \label{varycontour}
\end{figure}
We calculated the difference between the average of $B_\perp$ over the contour $C_+$ to the average over contour $C_-$. Averaged over all data sets, the difference between these two values is \SI{88}{\gauss}, which is equivalent to an uncertainty of $\pm \SI{44}{\gauss}$.  \\
Uncertainties in the $B_\perp$ maps have several sources. To begin with, the GRIS data (Stokes profiles) has an uncertainty. Optical properties along the beam path can also change over the years due to natural degradation and when readjusting optical elements. In addition, the seeing changes the profiles for each slit position, which translates into uncorrected straylight residuals. These errors are difficult to quantify. Furthermore, errors are introduced during our analysis. The SIR inversion code provides errors in terms of agreement between the fitted profiles and the input data, but these errors do not reflect how well the input model suits the complexity of the data. Our model assumes a magnetic field vector and a velocity that is constant in height. We chose this  simplification  to achieve robustness, so that the same inversion strategy can be used for all data sets. However, it does not capture the same features in the profiles that a more complex model would (e.g., three lobes in Stokes-\emph{V}). Further possible error sources for the $B_\perp$ or $B_\parallel$ value of single pixels are the azimuthal disambiguation and the transformation of the magnetic field vector from the LOS to the local reference frame. As we average over contours that typically consist of around \SIrange[range-units=single,range-phrase = -]{500}{1000}{} pixels, random errors are reduced for the quantities  $\overline{B_\perp}$ and $\overline{B_\parallel}$. \\
For this study we used the difference between the contours $C_+$ and  $C_-$ to derive an uncertainty of $\pm \SI{44}{\gauss}$ coming from manually identifying the intensity contours. As it is difficult to calculate an exact number for the uncertainty of the $B_\perp$ and $B_\parallel$ maps including  averaging over contours, we used a pragmatic approach by estimating that this uncertainty is comparable to the uncertainty in selecting the intensity contours. Conservatively rounding up, this leads to a total uncertainty of $\pm \SI{100}{\gauss}$ for $\overline{B_\perp}$ and $\overline{B_\parallel}$ values. This number is to be understood as an estimate rather than a precisely calculated quantity.

\subsection{Intensity and $B_\perp$ contours}
\label{geomdiff}
Another approach to check whether a constant value of $B_\perp$ can serve to distinguish between umbral and penumbral magneto-convection is to compare intensity contours to contours calculated in maps of $B_\perp$ at this constant value. For each data set we calculated contours in maps of $B_\perp$ at the average value of $\overline{B_\perp}=\SI{1787}{\gauss}$ calculated in section \ref{bversec}. In Fig. \ref{bvercont} these contours are compared to the manually identified intensity contours. We used the geometrical difference $\Delta P$ between these contours as a parameter to asses how well they coincide. As in \citetads{2018A&A...620A.104S}, we define $\Delta P$ as the area (in pixels) that is enclosed by one of the two contours, but not the other. This value is divided by the length of the contour (in pixels) for normalization. For each data set, this value is shown in the captions of the subplots in Fig. \ref{bvercont}. Additionally, we show a data set with a contour at $\overline{B_\mathrm{abs}}$ (average value of $B_\mathrm{abs}$ over the contours of all 11 sunspots). Like contours at average $\overline{B_\parallel}$ values, these contours do not outline the umbra. They are often fragmented and for large sunspots the contour lies outside the umbra--penumbra boundary, while it lies inside the umbra--penumbra boundary for smaller ones.

\begin{figure*}[h]
\centering
   \includegraphics[width=17cm]{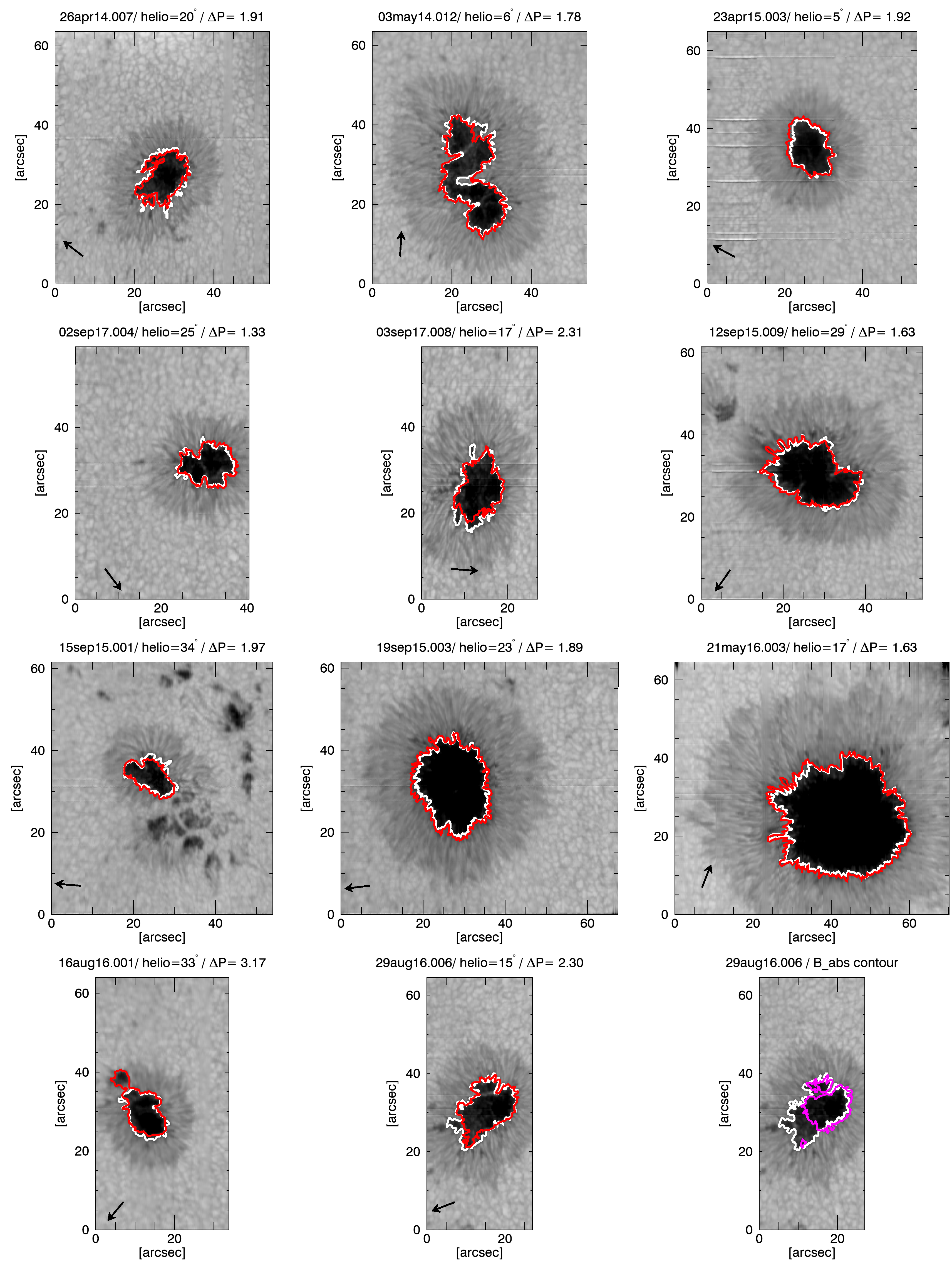}
     \caption{Continuum intensity maps of the sunspots observed with GRIS. White: Manually found intensity contours. Red: Contours calculated in $B_\perp$ maps at the value of \SI{1787}{\gauss} (average over all 11 sunspots). $\Delta$ P denotes the geometric difference between the two contours. The black arrow points towards disk center. The map in the bottom right corner shows an example of a contour using the average of the $B_{\mathrm{abs}}$ values in magenta.}
     \label{bvercont}
\end{figure*}


\section{Discussion}
\subsection{Interpreting $B_\perp$ averages}
We analyzed the vertical component of the magnetic field $B_\perp$ at the boundary between umbra and penumbra. The average value over the manually found intensity contour, $\overline{B_\perp}$, was compared to the average of the horizontal component,  $\overline{B_\parallel}$. The $\overline{B_\perp}$ values were less widely scattered between the data sets than the $\overline{B_\parallel}$ values (see Fig. \ref{bverrelsort}), resulting in a normalized standard deviation that is 2.4 times smaller. In addition, the absolute value of the standard deviation of $\overline{B_\perp}$ values for the different data sets was \SI{94}{\gauss}, which is in the same range as the estimated uncertainty of \SI{100}{\gauss}. We therefore interpret the deviations of $\overline{B_\perp}$ values from the average over the data sets as being caused by our measurement uncertainty. This result is in line with the finding of $B_\perp$ taking a constant value at the  umbra--penumbra boundary for stable sunspots \citepads{2018A&A...611L...4J}. \\
It should be noted, however, that there are single outliers in $B_\perp$ values along contours of single data sets (see Fig. \ref{alongcontour}). In Fig. \ref{bvercont} this is seen as regions where the contour at $B_\perp=\SI{1787}{\gauss}$ does not exactly match the intensity contours. This behavior is quantified as the geometric difference $\Delta P$. For all the data sets except 16aug16.001, this number is in the range of 2 pixels or smaller, so the deviations between the contours can be considered small. We interpret the locations where the contours do not match as regions where the umbra--penumbra boundary is not completely stable. A comparison with HMI continuum images accessed via JHelioviewer (not shown here) confirmed that the spots where the intensity contour and the contour at $B_\perp=\SI{1787}{\gauss}$ match least closely were indeed undergoing evolution during the time of the GRIS observations. The sunspot in data set 16aug16.001, for example, expelled the pore (enclosed by the red contour but not the white one) on subsequent days. In most regions, the intensity contours and the contours at $B_\perp=\SI{1787}{\gauss}$ match remarkably well. This is clearly not the case for contours plotted at the average value of $B_\parallel $ and $B_\mathrm{abs}$. In theses cases contours are often fragmented, and for large sunspots the contour lies outside the umbra--penumbra boundary, while it lies inside the umbra--penumbra boundary for smaller ones. An example of such a contour for one data set is shown in Fig. \ref{bvercont}. Overall, these findings support the existence of a constant value of $B_\perp$ at the umbra--penumbra boundary. \\

\subsection{Comparison to HMI and Hinode/SP values}
\label{comp_text}
In addition to providing further statistical evidence to the {Jur{\v{c}}{\'a}k criterion} from ground-based observations for the first time, we obtain the critical value of \SI[separate-uncertainty=true]{1787(100)}{\gauss} at the umbra--penumbra boundary for GRIS data in the \SI{1.5}{\micro \metre} (near-IR) range. Previously, this parameter had been determined only from Hinode/SP observations \citepads{2018A&A...611L...4J,2011A&A...531A.118J} and for HMI observations \citepads{2018A&A...620A.104S}. For the Hinode/SP data (Fe I \SI{6302}{\angstrom}) a value of $1867^{+18}_{-16}$ \SI{}{\gauss} was derived, and for HMI  (Fe I \SI{6173}{\angstrom}) a value of \SI[separate-uncertainty=true]{1692(15)}{\gauss}. The question is now how to interpret the differences of \SI{+80}{\gauss} (Hinode/SP) and \SI{-95}{\gauss} (HMI), respectively, from the GRIS value.
It should be noted that in all three studies a certain degree of arbitrariness is introduced by either setting a fixed intensity value to define the contours in  intensity \citepads{2018A&A...611L...4J,2018A&A...620A.104S} or by defining them manually as was done in this work. This effect is not included in the respective uncertainties of the HMI and Hinode values. As shown in Sect. \ref{intcont}, differences in $\overline{B_\perp}$ values in the \SIrange[range-units=single,range-phrase = -]{50}{100}{\gauss } range can be obtained by using a different intensity threshold. In this sense, all values are comparable. \\ 
In order to compare magnetic field values in general, both instrumental effects (e.g., the spectral and spatial resolution of the instrument, the seeing during the observations) and the atmospheric height ranges of the respective lines have to be considered. An overview of the parameters discussed in the following paragraphs is shown in Table \ref{line_properties}.

\begin{table}
\caption{Compilation of considered parameters used to  compare our result of $B_\perp^{\rm const}$ from GRIS to the previous results from SP/Hinode and SDO/HMI data: Line wavelength, spectral resolution, spatial resolution, straylight (residuals after correction), formation height difference (relative to GRIS), and the $B_\perp^{\rm const}$ value calculated in the respective studies. References to the displayed values are given in the text (Sect. \ref{comp_text}).}
\label{line_properties}
\centering 
\begin{tabular}{l  c c c }
Parameter & GRIS & SP/Hinode & SDO/HMI \\
\hline \hline 
$\lambda$ [\SI{}{\angstrom}]                        & $15648/15662$       & 6301/6302       & 6173 \\
spec. res. & \SI{113000}{} & \SI{150000}{} & \SI{44000}{} \\
spat. res. [$\arcsec$]     & $\sim$ 0.4                    & $\sim$ 0.64        & $\sim$ 1.0 \\
res. straylight                             & Yes                    & -                  & - \\
$\Delta$z\,[km]                               &        -           & $\sim$ 65                    & $\sim$ 65    \\
$B_\perp^{\rm const}$ [Gauss] & $(1787\pm100)$
& $1867^{+18}_{-16}$ 
& $ (1692\pm15) $ \\
\hline
\end{tabular}
 \end{table}

\paragraph{Formation ranges:} \citetads{2016A&A...596A...2B} derived that the Fe I \SI{15648}{\angstrom} lines are formed around \SIrange[range-units=single,range-phrase = -]{60}{70}{\kilo\metre} deeper than the Fe I  \SI{6302}{\angstrom} Hinode/SP lines. Given that in sunspots the vertical gradient of the magnetic field strength is \SIrange[range-units=single,range-phrase = -]{0.5}{3}{\gauss  / \kilo\metre}, as derived by \citetads{2006RPPh...69..563S}, we expect a difference in absolute magnetic field strength of  \SIrange[range-units=single,range-phrase = -]{30}{210}{\gauss}. Hitherto, the broad range of the gradient could not be narrowed down \citepads{2018SoPh..293..120B}. In our samples the average inclination value at the contours was $\approx$ \SI{35}{\degree}, leading to differences in $B_\perp$ of $\approx \SIrange[range-units=single,range-phrase = -]{25}{170}{\gauss}$. From this effect, a higher  $B_\perp$ would be expected for GRIS than for Hinode/SP. The same applies when comparing GRIS to HMI; the Fe I \SI{6302}{\angstrom} line (Hinode) and the Fe I \SI{6173}{\angstrom} line (HMI) have a comparable continuum opacity and the combination of oscillator strength and excitation potential is also similar. That leads to a comparable formation range.

\paragraph{Straylight:} In order to avoid an overcorrection of profiles, we applied a conservative spatial straylight correction to the GRIS data (see Sect. \ref{straysec}) and there is probably residual spatial straylight still left in the profiles. As the penumbra has higher intensity values than the umbra, pixels at the boundary are influenced more strongly by straylight from the penumbra than from the umbra. Although the lines are fully split, the peak of Stokes-V profiles, for example, can still be shifted towards the line center if a contribution from a neighboring profile with a smaller field strength value is added. Therefore, more straylight leads to lower field strength values at the umbra--penumbra boundary. For GRIS, lower $B_\perp$ values are expected than for Hinode/SP and HMI, which do not suffer from seeing effects.

\paragraph{Spatial resolution:} 
The spatial resolution of the respective instruments leads to a similar mixing of profiles from different spatial regions. Compared to the influence of seeing, the effect of different spatial resolution on $B_\perp$ values at the umbra--penumbra boundary is less dominant because the spatial dimensions are smaller. The values of GRIS ($\sim 0.4 \arcsec$ \citepads{2016A&A...596A...2B}) is lower than the values from Hinode/SP (fast maps, $\sim 0.64 \arcsec$ \citepads{2013SoPh..283..579L}) and SDO/HMI ($\sim 1 \arcsec$ \citepads{2014SoPh..289.3483H}). Although we expect the effect of difference in spatial resolution on magnetic field values to be small, it would lead to lower $B_\perp$ values at the umbra--penumbra boundary for lower spatial resolution. This is in accordance with the finding of \citetads{2017ApJ...851..111S}, who did a statistical analysis of differences between the magnetic field values from Hinode/SP and SDO/HMI. He found that, after the filling factor of the inversion and the spectral resolution, the spatial resolution was the parameter that had the smallest impact on differences in magnetic field values. 

\paragraph{Spectral resolution:} Another effect when comparing different data sources comes from the difference in spectral sampling and resolution. Inversion codes convolve synthetic profiles with the spectral point spread function (PSF) of the instrument before comparing to observed profiles. However, \citetads{2017ApJ...851..111S} argues that the spectral sampling is one of the dominant reasons for a systematic difference between HMI magnetic field values and Hinode magnetic field values they found in a statistical analysis. With this finding, the difference in $B_\perp$ between Hinode and HMI can be explained. Following this argument, higher field strength values should be obtained for higher spectral resolution. For the three instruments, the following values for the spectral resolution are given: GRIS \SI{113000}{} \citepads{2012AN....333..872C}, Hinode \SI{150000}{}, HMI \SI{44000}{} (values for Hinode and HMI are derived from the spectral sampling given in \citetads{2017ApJ...851..111S}). For GRIS, an intermediate value between HMI and Hinode should then be expected for $B_{\perp}$.
\\ \\

In summary, from the difference in formation height, a higher value for the average $B_\perp$ from GRIS data would be expected in comparison to Hinode and HMI, but the residual spatial straylight leads to a reduction in the value. Together with the spectral sampling, this can explain why we measure a value that is between the value from Hinode/SP and HMI. However, a quantification of these effects is not possible because the difference due to formation heights has a high uncertainty. For the differences due to spatial straylight and spectral resolution, the authors are not aware of quantitative numbers for comparing ground-based to space-based spectropolarimetric solar data. \\

\section{Conclusion}
 Our study provides additional proof for the {Jur{\v{c}}{\'a}k criterion} ($B_\perp$ reaches a constant value $B_\perp^{\rm const}$ at the umbra--penumbra boundary) for ground-based data in the infrared regime. The data sources HMI, Hinode/SP, and GRIS have different data characteristics (ground-based versus space-based, different spatial and spectral resolutions, different formation heights, etc.), and therefore the measured value of $B_\perp^{\rm const}$ is different within a plausible range of less than \SI{200}{\gauss}. We conclude that the three studies for the different instruments show that stable sunspots have the fundamental property of an absolute constant value of $B_\perp$ at the umbra--penumbra boundary. This empirical finding can help to verify analysis methods that calculate magnetic field parameters, especially inversions. The value of  $B_\perp$ should be close to constant at the umbra--penumbra boundary, and if $B_\perp^{\rm const}$ has already been calculated for respective data this average value should be met. This property also serves as a requirement for numerical simulations of sunspots to be realistic.  \\ \\
 Another application of the  {Jur{\v{c}}{\'a}k criterion} is to judge whether or not the umbra--penumbra boundary is stable in a specific region. Unstable regions are identified as regions where the umbra--penumbra boundary, as seen in intensity and the (iso-)contour at $B_\perp=B_\perp^{\rm const}$, do not match (see Fig. \ref{bvercont}). Such unstable regions are expected to be undergoing evolution. The sunspot could, for example, still be in the process of forming its penumbra (e.g., \citetads{2014PASJ...66S...3J,2017A&A...597A..60J}); it could be about to expel a pore from the umbra; or it could be about to enter the decaying phase. An example of a decaying sunspot, where $B_\perp$ is not   constant, is also given by \citetads{2018A&A...620A.191B}. The geometric difference, $\Delta P$, between the intensity contour and the contour at $B_\perp = B_\perp^{\rm const}$ (see Sect. \ref{geomdiff} and \citetads{2018A&A...620A.104S}) quantifies the stability over the whole contour into one number for each sunspot. The value of $\Delta P$ is small for stable sunspots and larger for sunspots under evolution, and therefore acts as an index of stability. This method has the potential to provide indications about the future and past evolution of a sunspot from only one snapshot. More statistics, however, are needed to quantitatively assess the reliability of such predictions. \\ \\
In addition, further studies are needed  not only to investigate sunspots, but also to focus on pores and whether the boundary between the pore and the surrounding quiet sun possesses a similar property (see \citet{marta}). Stable pores should not include any regions with  $B_\perp <B_\perp^{\rm const}$. Expanding the  {Jur{\v{c}}{\'a}k criterion} to a more general criterion on whether convection is suppressed, further studies could also examine other magneto-convective modes, for example    umbral dots. The aim would be to check whether a similar value for $B_\perp$ governs the suppression or admission of convection for these cases. 
\\


\begin{acknowledgements} 
The 1.5-meter GREGOR  solar telescope was built by a German consortium under the leadership of the Kiepenheuer Institut f\"ur Sonnenphysik in Freiburg with the Leibniz Institut f\"ur Astrophysik Potsdam, the Institut f\"ur Astrophysik G\"ottingen, and the Max-Planck Institut f\"ur Sonnensystemforschung in G\"ottingen as partners, and with contributions by the Instituto de Astrof\'sica de Canarias and the Astronomical Institute of the Academy of Sciences of the Czech Republic. The GRIS instrument was developed thanks to the support by the Spanish Ministry of Economy and Competitiveness through the project AYA2010-18029 (Solar Magnetism and Astrophysical Spectropolarimetry). \\ \\
We would like to thank Juan Manuel Borrero for his great help with the SIR inversions, Svetlana Berdyugina for the enlightening discussions and Reza Rezaei for his help in setting up the AZAM code. We also want to thank the Science Date Centre team at Leibniz-Insititut für Sonnenphysik (KIS) for providing and organizing a large amount of data openly to the community  (\href{sdc.leibniz-kis.de}{sdc.leibniz-kis.de}) and especially Morten Franz for his help with working on the data. \\
We would also like to thank the anonymous referee for valuable contributions and constructive comments.

\end{acknowledgements}

\bibliographystyle{aa} 
\bibliography{bvergris_paper} 

\end{document}